\documentclass[showpacs,aps,prd,preprint,nofootinbib,showkeys,unsortedaddress,raggedbottom]{revtex4-1}
\pdfoutput=1

\usepackage{bm}
\usepackage{amsmath}
\usepackage{graphicx}
\usepackage{subfigure}
\usepackage[usenames,dvipsnames]{color}
\definecolor{darkblue}{RGB}{0,0,196}
\usepackage[colorlinks=true,linkcolor=darkblue,citecolor=darkblue,urlcolor=darkblue]{hyperref}

\usepackage{setspace}
\usepackage{footmisc}
\usepackage[makeroom]{cancel}

\newcommand{\intdP}{\int\!dP}

\def\be{\begin{equation}}
\def\ee{\end{equation}}
\def\ba{\begin{eqnarray}}
\def\ea{\end{eqnarray}}

\begin{document}

\title{Anisotropic hydrodynamic modeling of 2.76 TeV Pb-Pb collisions}

\author{Mubarak Alqahtani} 

\author{Mohammad Nopoush} 
\affiliation{Department of Physics, Kent State University, Kent, OH 44242 United States}

\author{Radoslaw Ryblewski}

\affiliation{Institute of Nuclear Physics, Polish Academy of Sciences, PL-31342 Krak\'ow, Poland}

\author{Michael Strickland} 

\affiliation{Department of Physics, Kent State University, Kent, OH 44242 United States}

\begin{abstract}
We compare phenomenological results from 3+1d quasiparticle anisotropic hydrodynamics (aHydroQP)  with experimental data collected in LHC 2.76 TeV Pb-Pb collisions. In particular, we present comparisons of particle spectra, average transverse momentum, elliptic flow, and HBT radii. The aHydroQP model relies on the introduction of a single temperature-dependent quasiparticle mass which is fit to lattice QCD data. By taking moments of the resulting Boltzmann equation, we obtain the dynamical equations used in the hydrodynamic stage which include the effects of both shear and bulk viscosities. At freeze-out, we use anisotropic Cooper-Frye freeze-out performed on a fixed-energy-density hypersurface to convert to hadrons. To model the production and decays of the hadrons we use  THERMINATOR 2 which is customized to sample from ellipsoidal momentum-space distribution functions. Using smooth Glauber initial conditions, we find very good agreement with many heavy-ion collision observables. 
\end{abstract}

\date{\today}

\pacs{12.38.Mh, 24.10.Nz, 25.75.Ld, 47.75.+f, 31.15.xm}

\keywords{Quark-gluon plasma, Relativistic heavy-ion collisions, Anisotropic hydrodynamics, Equation of state, Quasiparticle, Boltzmann equation}

\maketitle

\section{Introduction}
\label{sec:intro}

Relativistic hydrodynamics has been quite successful in describing the soft hadron spectra ($p_T \lesssim 2$ GeV) and collective flow observed in ultrarelativistic heavy-ion collision (URHIC) experiments at the Relativistic Heavy Ion Collider (RHIC) and Large Hadron Collider (LHC) \cite{Huovinen:2001cy,Hirano:2002ds,Kolb:2003dz,Muronga:2001zk, Muronga:2003ta, Muronga:2004sf, Heinz:2005bw, Baier:2006um, Romatschke:2007mq, Song:2007fn, Dusling:2007gi, Song:2007ux, Baier:2007ix, Luzum:2008cw, Song:2008hj, Broniowski:2008qk, Heinz:2009xj, Denicol:2010tr, Denicol:2010xn, Schenke:2010rr, Bozek:2010vz,Schenke:2011tv, Bozek:2011wa, Niemi:2011ix, Niemi:2012ry, Bozek:2012qs, Bozek:2012en,Denicol:2012cn,Denicol:2014vaa,Ryu:2015vwa,Tinti:2016bav} (see \cite{Heinz:2013th, Gale:2013da, Jeon:2016uym,Jaiswal:2016hex} for recent reviews).  Early on, the success of this program suggested that the quark-gluon plasma (QGP) generated in URHICs was nearly isotropic in the local rest frame (LRF); however, in practice, one finds that there are rather large momentum-space anisotropies, which are driven primarily by the rapid longitudinal expansion of the QGP created in URHICs \cite{Ryblewski:2013jsa,Strickland:2014pga}.  All studies indicate that at early times after the nuclear impact, the QGP possesses a high degree of momentum-space anisotropy in the fluid LRF, ${\cal P}_T/{\cal P}_L \gg 1$, which only slowly relaxes towards unity during the QGP liftetime.  Additionally, at all proper times, there are large momentum-space anisotropies near the transverse/longitudinal ``edges'' of the fireball where the system is nearly free streaming.  The situation only gets worse as one goes from AA to pA and pp collisions, since gradients are larger and system lifetimes are considerably shorter.  As a consequence, larger non-equilibrium deviations are expected for these systems, which push traditional viscous hydrodynamics to its limits~\cite{Nopoush:2015yga}.  This has motivated the investigation of alternative formulations of dissipative relativistic hydrodynamics which can be applied to systems which might possess a high degree of momentum-space anisotropy at all points in spacetime~\cite{Ryblewski:2013jsa,Strickland:2014pga}.

One way to proceed is to reorganize the expansion of the one-particle distribution function around a leading-order form which possesses intrinsic momentum-space anisotropies but still guarantees positivity~\cite{Florkowski:2010cf,Martinez:2010sc}.  This method has become known as anisotropic hydrodynamics (aHydro) and there are now many groups pursuing this idea~\cite{Ryblewski:2010ch,Martinez:2012tu,Ryblewski:2012rr,Bazow:2013ifa,Tinti:2013vba,Nopoush:2014pfa,Tinti:2015xwa,Bazow:2015cha,Strickland:2015utc,Alqahtani:2015qja,Molnar:2016vvu,Molnar:2016gwq,Alqahtani:2016rth,Bluhm:2015raa,Bluhm:2015bzi}.  One of the selling points for the aHydro approach has been that it better reproduces exact solutions to the Boltzmann equation compared to traditional near-equilibrium viscous hydrodynamic approaches, even in the limit of very large shear viscosity to entropy density ratio and/or initial momentum-space anisotropy \cite{Florkowski:2013lza,Florkowski:2013lya,Denicol:2014tha,Denicol:2014xca,Nopoush:2014qba,Heinz:2015gka,Molnar:2016gwq,Martinez:2017ibh}.  Given this success, the focus has recently turned to making aHydro a practical phenomenological tool with a realistic equation of state (EoS) and self-consistent anisotropic hadronic freeze-out.  In a previous short paper~\cite{Alqahtani:2017jwl}, we presented the first comparisons of experimental data with phenomenological results obtained using generalized 3+1d aHydro including:  (1) three momentum-space anisotropy parameters in the underlying distribution function, (2) the quasiparticle aHydro (aHydroQP) method for implementing a realistic EoS \cite{Alqahtani:2015qja, Alqahtani:2016rth, Alqahtani:2016ayv}, and (3) anisotropic Cooper-Frye freeze-out~\cite{Nopoush:2015yga,Alqahtani:2016rth} using the same form for the distribution function as was assumed for the dynamical equations.  

To the best of our knowledge, all previous phenomenological applications of aHydro have either relied on the approximate conformal factorization of the energy-momentum tensor, see e.g.~\cite{Martinez:2012tu,Ryblewski:2012rr,Nopoush:2016qas,Strickland:2016ezq}, and/or have used isotropic freeze-out \cite{Ryblewski:2012rr}.  To address the issue with freeze-out, we use a customized version of THERMINATOR 2~\cite{Chojnacki:2011hb} which has been modified to accept ellipsoidally-anisotropic distribution functions.  In this paper, we extend our previous work \cite{Alqahtani:2017jwl} by (1) giving the details of the formalism of 3+1d aHydroQP and (2) presenting comparisons with a wider set of heavy-ion observables. Here we present comparisons of charged-hadron multiplicity, identified-particle spectra, identified-particle average transverse momentum, charged-particle elliptic flow, identified-particle elliptic flow, the integrated elliptic flow vs pseudorapidity, and the HBT radii. For some observables, such as  the spectra, compared to Ref.~\cite{Alqahtani:2017jwl} we present comparisons in more centrality classes and with higher statistics.
	
The structure of the paper is as follows. In Sec.~\ref{sec:setup}, we present our general setup. In Sec.~\ref{sec:equations}, we  introduce the formalism for quasiparticle anisotropic hydrodynamics and then derive  the 3+1d dynamical equations. In Sec.~\ref{sec:freeze-out}, we discuss anisotropic Cooper-Frye freeze-out. In Sec.~\ref{sec:results}, we compare our model results obtained using the 3+1d aHydroQP model for Pb-Pb collisions at LHC energies with data from the ALICE collaboration. Sec.~\ref{sec:conclusions} contains our conclusions and an outlook for the future. In App.~\ref{app:identities}, we list the derivatives used in the body of the paper. App.~\ref{app:h-functions} contains details concerning the thermodynamic integrals introduced in the body of the text and our optimized scheme for their evaluation.

\section{Setup}
\label{sec:setup}
\subsection{Conventions and notation}
\label{subsec:notations}
Here  we define some conventions and notation that will be used in the body of this paper. The metric is taken to be ``mostly minus'' with $x^\mu = (t, x, y, z)$  where the line element is $ds^2=g_{\mu\nu} dx^\mu dx^\nu=dt^2-dx^2-dy^2-dz^2$ with $g^{\mu\nu}$ being metric tensor in Minkowski space. The longitudinal proper time is $ \tau = \sqrt{t^2-z^2}$ and the longitudinal spacetime rapidity is $\varsigma = \tanh^{-1} (z/t)$.

The basis vectors for a general 3+1d system in the laboratory frame can be obtained by the following parametrization, which is based on a set of Lorentz transformations applied to the LRF basis vectors \cite{ Martinez:2012tu, Ryblewski:2010ch}. The set of successive transformations correspond to a longitudinal boost $\vartheta$ along the beam line, a rotation $\varphi\equiv \tan^{-1}(u_y/u_x)$ around the beam line, and a transverse boost $\theta_\perp$, which together yield
\ba
u^\mu &\equiv& (u_0 \cosh\vartheta,u_x,u_y,u_0 \sinh\vartheta) \, , \nonumber\\
X^\mu &\equiv& \Big(u_\perp\cosh\vartheta,\frac{u_0 u_x}{u_\perp},\frac{u_0 u_y}{u_\perp},u_\perp\sinh\vartheta\Big) , \nonumber \\ 
Y^\mu &\equiv& \Big(0,-\frac{u_y}{u_\perp},\frac{u_x}{u_\perp},0\Big)  , \nonumber \\
Z^\mu &\equiv& (\sinh\vartheta,0,0,\cosh\vartheta ) \, ,
\label{eq:4vectors}
\ea
where $u^\mu$ is the fluid four-velocity and $u_\perp\equiv \sqrt{u_x^2+u_y^2}=\sqrt{u_0^2-1} = \sinh\theta_\perp$.

\subsection{Distribution Function}
\label{subsec:distribution}

The leading order distribution function in aHydro is assumed to be of generalized Romatschke-Strickland form \cite{Romatschke:2003ms,Romatschke:2004jh}
\be
f(x,p) = f_{\rm eq}\!\left(\frac{1}{\lambda}\sqrt{p_\mu \Xi^{\mu\nu} p_\nu}\right) ,
\label{eq:genf}
\ee
where $\lambda$ is an energy scale which becomes the temperature in the isotropic equilibrium limit. The anisotropy tensor has the form $\Xi^{\mu\nu} \equiv u^\mu u^\nu + \xi^{\mu\nu} - \Delta^{\mu\nu} \Phi$ where $\xi^{\mu \nu}$ is a symmetric traceless tensor obeying $ u_\mu \xi ^{\mu \nu} = 0$ and $ {\xi^\mu}_\mu = 0 $, $\Phi$ is the bulk degree of freedom, and $\Delta^{\mu\nu} = g^{\mu\nu} - u^\mu u^\nu$ is the transverse projector \cite{ Martinez:2012tu, Nopoush:2014pfa}. Using the ellipsoidal form (\ref{eq:genf}) and the tracelessness of $\xi^{\mu\nu}$, we are left with three independent parameters out of the four original parameters $\Phi$ and ${\boldsymbol \xi} = (\xi_x,\xi_y,\xi_z)$. In thermal equilibrium, the distribution function $f_{\rm eq}(x)$ can be identified as Fermi-Dirac, Bose-Einstein, or Maxwellian distribution. Herein, ignoring the quantum statistics, we take the Boltzmann form with zero chemical potential. 

We note that, in order to perform the integrals, it is useful to define three parameters $\alpha_i$ as the system's independent anisotropy parameters
\be
\alpha_i \equiv (1 + \xi_i + \Phi)^{-1/2} \, ,
\label{eq:alphadef}
\ee  
such that the distribution function can be written as
\be 
f(x,p)= f_{\rm eq}\!\left(\frac{1}{\lambda}\sqrt{\sum_i \frac{p_i^2}{\alpha_i^2}+m^2}\right) .
\label{eq:genf2}
\ee

\subsection{The equation of state}
\label{subsec:EoS}

We use an analytic parameterization of lattice QCD (LQCD) data for the trace anomaly taken from the Wuppertal-Budapest collaboration \cite{Borsanyi:2010cj}.  This analytic parameterization can be used to calculate the energy density, pressure, and entropy density using standard thermodynamic identities.  In order to implement the equation of state in the quasiparticle model, we fit a single temperature-dependent mass, $m(T)$, to the LQCD entropy density.  For details, we refer the reader to  Ref.~\cite{Alqahtani:2015qja}.

\subsection{Shear and bulk viscosity in the quasiparticle (QP) model}
\label{subsec:shear}

The relaxation time $\tau_{\rm eq}$ can be related to the shear viscosity, $\eta$, for the system of quasiparticles.  The shear viscosity for a quasiparticle gas can be found in Refs.~\cite{Romatschke:2011qp} and \cite{Tinti:2016bav}, Eqs.~(4.3) and (46), respectively, with both giving the same result
\be 
\frac{\eta}{\tau_{\rm eq}} = \frac{1}{T} I_{3,2}(\hat{m}_{\rm eq}) \,\, ,
\label{eq:eta}
\ee
with $\hat{m}_{\rm eq} \equiv m/T$ and 
\ba 
I_{3,2}(x) &=& \frac{N_{\rm dof} T^5\, x^5}{30 \pi^2} \bigg[ \frac{1}{16} \Big(K_5(x)-7K_3(x)+22 K_1(x) \Big)-K_{i,1}(x) \bigg]  \, , \nonumber \\ 
K_{i,1}(x)&=&\frac{\pi}{2}\Big[1-x K_0(x) {\cal S}_{-1}(x)-xK_1(x) {\cal S}_0(x)\Big] \, ,
\ea
where $N_{\rm dof}$ is the number of degrees of freedom, $K_n$ are modified Bessel functions of the second kind, and ${\cal S}_n$ are modified Struve functions.
Using Eq.~(\ref{eq:eta}), the relaxation time can be written as 
\ba 
\tau_{\rm eq}(T)= \bar{\eta} \, \frac{{\cal E+P}}{I_{3,2}(\hat{m}_{\rm eq})} \, ,
\ea
where $\bar{\eta} \equiv \eta/s$ with $s$ being the entropy density, ${\cal E}$ is the energy density, and ${\cal P}$ is the pressure.

\begin{figure}[t!]
\vspace{3.5mm}
\includegraphics[width=.99\linewidth]{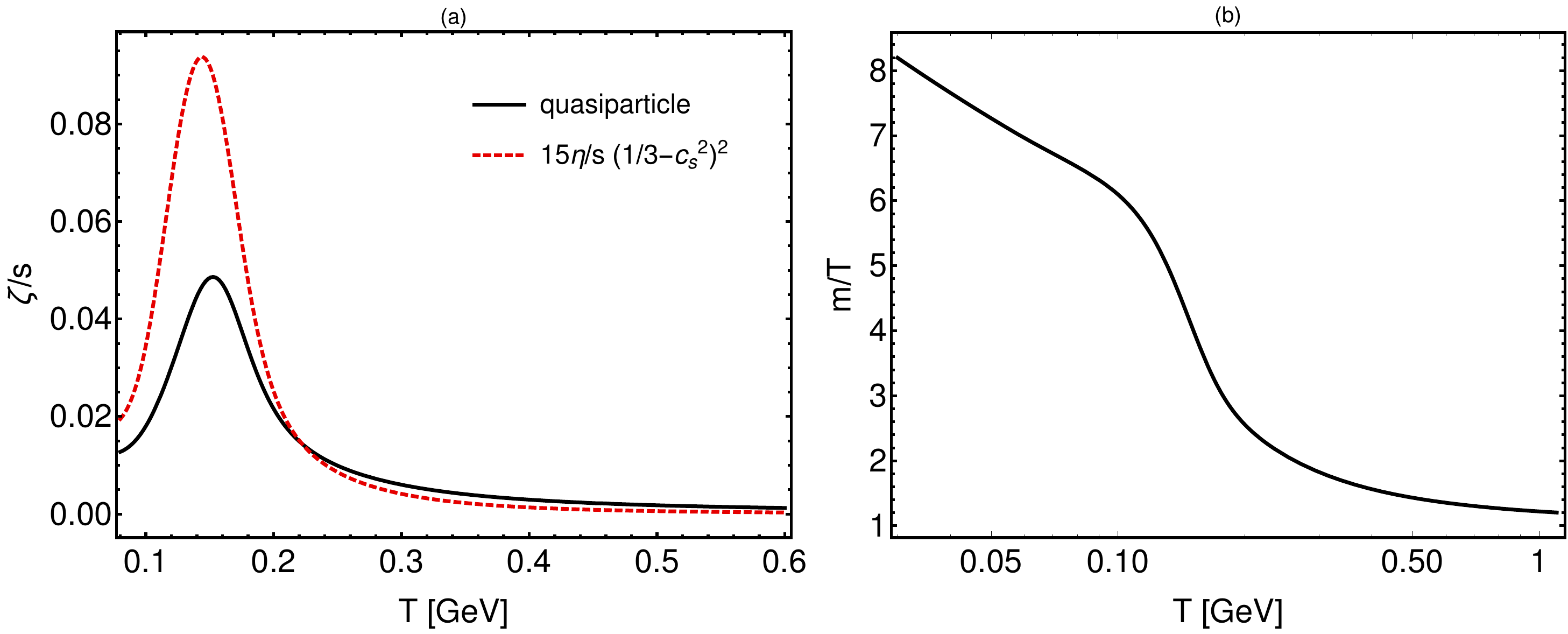}
\caption{In panel (a), we show the bulk viscosity scaled by the entropy density obtained using the quasiparticle model (black solid line) \cite{Romatschke:2011qp,Tinti:2016bav}  compared with $\zeta/s = 15  \eta/s \, (1/3 - c_s^2)^2$, which is a frequently used small-mass limit result (red dashed line).  In panel (b), we show $m/T$ extracted by fitting to LQCD results for the entropy density \cite{Borsanyi:2010cj}.}
\label{fig:zeta}
\end{figure}

Using the quasiparticle model, one can extract the bulk viscosity in a similar manner.  Expressions for the bulk viscosity for a quasiparticle gas can be found in Refs.~\cite{Romatschke:2011qp} and \cite{Tinti:2016bav}, Eqs.~(4.4) and (45), respectively, with, both again giving the same result
\be 
\frac{\zeta}{\tau_{\rm eq}} = 
\frac{5}{3 T} I_{3,2} (\hat{m}_{\rm eq})
- c_s^2 ({\cal E} +{\cal P}) 
+ c_s^2 m \frac{dm}{dT} I_{1,1}(\hat{m}_{\rm eq}) \, ,\ee
where 
\be
I_{1,1}(x) = \frac{N_{\rm dof} T^3 \, x^3}{6 \pi^2} \Bigg[ \frac{1}{4} \Big( K_3(x)-5K_1(x)\Big)+K_{i,1} (x)\Bigg] \, .   
\ee
In Fig.~\ref{fig:zeta}-a, we plot the result for the bulk viscosity to entropy density ratio $\zeta/s$ in the quasiparticle model as a black solid line. For comparison, we plot  \mbox{$\zeta/s = 15 \eta/s \, (1/3 - c_s^2)^2$}, where $c_s$ is the speed of sound.  This is an often-used small-mass expansion result \cite{Weinberg:1972}. For both curves, we assume that $\eta/s = 2/(4\pi)$.

In Fig.~\ref{fig:zeta}-b we plot $m/T$ obtained by fitting to the Wuppertal-Budapest LQCD results for the entropy density~\cite{Alqahtani:2015qja}.  We point out that, at small temperatures, the value of $m/T$ necessary to fit the LQCD data \cite{Borsanyi:2010cj} is not small, invalidating the often used small $m/T$ expansion used to compute effective transport coefficients. The quasiparticle model has a finite bulk viscosity to entropy density ratio as shown in Fig.~\ref{fig:zeta}-a. It peaks in the vicinity of the phase transition temperature from QGP to a hadronic gas. By comparing to other ans\"atze for $\zeta/s$ used in other studies we see that, in our quasiparticle model, the peak value is much smaller $\sim 0.05$, compared to prior works.  For example, in Ref.~\cite{Ryu:2015vwa} the respective peak value of $\zeta/s$ is approximately $0.3$.

\section{dynamical equations}
\label{sec:equations} 

For a system of quasiparticles with a temperature-dependent mass, the Boltzmann equation is \cite{Jeon:1995zm,Romatschke:2011qp,Alqahtani:2015qja}
\ba
p^\mu\partial_\mu f+\frac{1}{2}\partial_i m^2\partial^i_{(p)}f =-{\cal C}[f]\,,
\ea
with $i \in \{x,y,z\}$.  Herein, we take the collisional kernel ${\cal C}[f]$ in the relaxation time approximation, $C[f] = p_\mu u^\mu ( f - f_{\rm eq})/\tau_{\rm eq}$.  By taking moments of this equation we can generate the necessary dynamical equations.

The kinetic part of the energy-momentum tensor can be obtained from the second moment of distribution function, however, when the quasiparticle mass is temperature dependent, this quantity is not conserved by itself.  In order to enforce thermodynamic consistency and energy-momentum conservation, one must introduce the background field contribution \cite{gorenstein1995gluon}, $B(T)$, which is fixed through comparison with LQCD data \cite{Romatschke:2011qp,Alqahtani:2015qja}
\ba
T^{\mu\nu}=T^{\mu\nu}_{\rm kinetic}+B(T) g^{\mu\nu} \, ,
\ea
where $T^{\mu\nu}_{\rm kinetic} = \int dP \, p^\mu p^\nu f(x,p)$ with $dP \equiv E^{-1}\,d^3p/(2\pi)^3$ being the Lorentz-invariant integration measure.

For the case considered here, namely a diagonal anisotropy tensor, the full energy-momentum tensor can be expanded as
\be
T^{\mu\nu}={\cal E}u^\mu u^\nu+{\cal P}_x X^\mu X^\nu+{\cal P}_y Y^\mu Y^\nu+{\cal P}_z Z^\mu Z^\nu \, .
\label{eq:T-expan}
\ee
The resulting energy density and pressures are
\ba
{\cal E} &=& {\cal H}_{3}({\boldsymbol\alpha},\hat{m}) \, \lambda^4+B \, ,\nonumber \\
{\cal P}_i &=& {\cal H}_{3i}({\boldsymbol\alpha},\hat{m}) \, \lambda^4-B \, ,
\ea
with $ i \in \{x,y,z\}$ and the ${\cal H}$-functions appearing above defined in App.~\ref{app:h-functions}.

\noindent
Taking the first moment of Boltzmann equation we obtain four equations
\ba
D_u{\cal E}+{\cal E}\theta_u+ {\cal P}_x u_\mu D_xX^\mu+ {\cal P}_y u_\mu D_yY^\mu +{\cal P}_z u_\mu D_zZ^\mu &=&0\, , \nonumber\\
D_x {\cal P}_x+{\cal P}_x\theta_x -{\cal E}X_\mu D_uu^\mu -{\cal P}_y X_\mu D_yY^\mu - {\cal P}_z X_\mu D_zZ^\mu &=& 0\,, \nonumber\\
D_y {\cal P}_y+{\cal P}_y \theta_y-{\cal E}Y_\mu D_uu^\mu -{\cal P}_x Y_\mu D_xX^\mu - {\cal P}_z Y_\mu D_zZ^\mu  &=& 0\,, \nonumber\\
D_z {\cal P}_z+{\cal P}_z \theta_z-{\cal E}Z_\mu D_uu^\mu- {\cal P}_x Z_\mu D_xX^\mu - {\cal P}_y Z_\mu D_yY^\mu &=& 0\,.
\label{eq:1stmoment}
\ea

\noindent
The second moment of Boltzmann equation involves a rank-3 tensor ${\cal I}$ which is the third moment of the distribution function
\be
{\cal I}^{\mu\nu\lambda} \equiv \intdP \, p^\mu p^\nu p^\lambda  f(x,p) \, .
\label{eq:I-int}
\ee
Expanding ${\cal I}$ over the basis vectors one has
\ba
{\cal I} &=&\, {\cal I}_u \left[ u\otimes u \otimes u\right] 
\nonumber \\
&+&\, {\cal I}_x \left[ u\otimes X \otimes X +X\otimes u \otimes X + X\otimes X \otimes u\right]
+(X\rightarrow Y) +(X\rightarrow Z)  \, ,
 \label{eq:Theta}
\ea
with \cite{Nopoush:2014pfa}
\ba
{\cal I}_i &=& \alpha \, \alpha_i^2 \, {\cal I}_{\rm eq}(\lambda,m) \, , \nonumber \\ 
{\cal I}_{\rm eq}(\lambda,m) &=&  4 \pi {\tilde N} \lambda^5 \hat{m}^3 K_3(\hat{m}) \, ,
\ea
where $\alpha = \prod_i \alpha_i$.  Since we have  eight independent variables, $\alpha_x$, $\alpha_y$, $\alpha_z$, $u_x$, $u_x$, $\vartheta$, $ T$, and $\lambda$, we need eight equations to solve the full 3+1d aHydroQP system.   With ten equations obtained from the second moment of Boltzmann equation, the system is overdetermined. Therefore, we have to come up with a selection rule to choose a subset of these equations in order to close the set of dynamical equations. The final equations are taken from the three diagonal projections of the equation of motion of the third moment, $X_\mu X_\nu \partial_\alpha {\cal I}^{\alpha \mu \nu}$, $Y_\mu Y_\nu \partial_\alpha {\cal I}^{\alpha \mu \nu}$, and $Z_\mu Z_\nu \partial_\alpha {\cal I}^{\alpha \mu \nu}$ giving~\cite{Alqahtani:2015qja}
\ba
D_u {\cal I}_x + {\cal I}_x (\theta_u + 2 u_\mu D_x X^\mu)
&=& \frac{1}{\tau_{\rm eq}} \Big[ {\cal I}_{\rm eq}(T,m) - {\cal I}_x \Big] \, ,  \nonumber \\
D_u {\cal I}_y + {\cal I}_y (\theta_u + 2 u_\mu D_y Y^\mu)
&=& \frac{1}{\tau_{\rm eq}} \Big[ {\cal I}_{\rm eq}(T,m) - {\cal I}_y \Big] \, , \nonumber\\
D_u {\cal I}_z + {\cal I}_z (\theta_u + 2 u_\mu D_z Z^\mu)
&=& \frac{1}{\tau_{\rm eq}} \Big[ {\cal I}_{\rm eq}(T,m) - {\cal I}_z \Big] \, . \label{eq:2ndmoment} 
\ea
Finally, in order to compute the local effective temperature $T$, we match the non-equilibrium kinetic energy density with the equilibrium kinetic energy density
\be
{\cal H}_3({\boldsymbol\alpha},\hat{m}) \lambda^4 = {\cal H}_{3,\rm eq}(1,\hat{m}_{\rm eq}) T^4.
 \label{eq:matching}
\ee
To summarize, to perform the numerical simulations reported herein we use the eight equations resulting from the first  (\ref{eq:1stmoment})  and second (\ref{eq:2ndmoment})  moments of the Boltzmann equation, together with the matching condition (\ref{eq:matching}).
 

\section{Anisotropic freeze-out}
\label{sec:freeze-out}

The QGP undergoes freeze-out at late times/low temperatures and the degrees of freedom need to be changed from hydrodynamical variables to hadronic positions and momenta. In this work, we perform ``anisotropic Cooper-Frye freeze-out'' using Eq.~(\ref{eq:genf2}) as the form for the one-particle distribution function. The anisotropic distribution function used in the freeze-out is guaranteed to be positive-definite, by construction, in all regions in phase space, avoiding the usual problems encountered within standard viscous hydrodynamic freeze-out. In practice, we construct a constant energy-density hypersurface, defined through $T_{\rm FO}={\cal E}^{-1}({\cal E}_{\rm FO})$. Then, by computing the number of particles that cross this hypersurface, one can determine the number of hadrons produced in heavy-ion collisions at freeze-out using
\begin{equation}
\bigg(p^0\frac{dN}{d^3p}\bigg)_{i}=\frac{{\cal N}_i}{(2\pi)^3}\int \! f_i(x,p) \, p^\mu d^3\Sigma_\mu \, ,
\label{eq:dNdp3}
\end{equation}
where $i$ labels the hadronic species, ${\cal N}_i\equiv 2s_i+1 $ is the degeneracy factor with $s_i$ being the spin of particle species $i$, and $f_i$ is the distribution function for particle species $i$ taking into account the appropriate quantum statistics.\footnote{In THERMINATOR 2, different isospin states are treated separately negating the need for an explicit isospin degeneracy factor.} For more details we refer the reader to \cite{Nopoush:2015yga,Alqahtani:2016rth}.

\section{Numerical results}
\label{sec:results}

In this section, we present  comparisons of our aHydroQP model  results with $\sqrt{s_{NN}}$ = 2.76 TeV Pb-Pb collision data available from the ALICE collaboration.  To set the initial conditions, we assume the system to be initially isotropic in momentum space ($ \alpha_{i}(\tau_0)=1 $), with zero transverse flow ($ {\bf u}_{\perp}(\tau_0) =0$), and Bjorken flow in the longitudinal direction ($ \vartheta(\tau_0) = \eta $). The initial energy density distribution in the transverse plane is computed from a ``tilted'' profile \cite{Bozek:2010bi}.  The distribution used is a linear combination of smooth Glauber wounded-nucleon and binary-collision density profiles, with a binary-collision mixing factor of $\chi = 0.15$.  In the longitudinal direction, we used a profile with a central plateau and Gaussian ``tails'', resulting in a longitudinal profile function of the form 
\be
\rho(\varsigma) \equiv \exp \left[ - (\varsigma - \Delta \varsigma)^2/(2 \sigma_\varsigma^2) \, \Theta (|\varsigma| - \Delta \varsigma) \right] \, .
\label{eq:rhofunc}
\ee
The parameters entering (\ref{eq:rhofunc}) were fitted to the pseudorapidity distribution of charged hadrons with the results being $\Delta\varsigma = 2.3$ and $\sigma_{\varsigma} = 1.6$.  The first quantity sets the width of the central plateau and the second sets the width of the Gaussian ``tails''.  

The resulting initial energy density at a given transverse position ${\bf x}_\perp$ and spatial rapidity $\varsigma$ was computed using 
\be 
{\cal E}({\bf x}_\perp,\varsigma) \propto (1-\chi) \rho(\varsigma) \Big[ W_A({\bf x}_\perp) g(\varsigma) + W_B({\bf x}_\perp) g(-\varsigma)\Big] + \chi \rho(\varsigma) C({\bf x}_\perp) \, ,
\ee
 where $W_{A,B}({\bf x}_\perp)$ is the wounded nucleon density for nucleus $A$ or $B$, $C({\bf x}_\perp)$ is the binary collision density, and $g(\varsigma)$ is the ``tilt function''.  The tilt function  is defined through
\ba 
g(\varsigma) =
\left\{ \begin{array}{lcccc}
0  & \,\,\,\,\,\,\,\,\,\,\,\,\,\,\,\ & \mbox{if} & \,\,\,
& \varsigma < -y_N \, ,
 \\ (\varsigma+y_N)/(2y_N) & & \mbox{if} &
& -y_N \leq \varsigma \leq y_N \, , \\
1 & & \mbox{if} & 
& \varsigma > y_N\, ,
\end{array}\right. \,\,\,\,\,\,\,\,\,\,\,\,\,
\ea 
where $y_N = \log(2\sqrt{s_{NN}}/(m_p + m_n))$ is the nucleon momentum rapidity \cite{Bozek:2010bi}.

\begin{figure}[t!]
\includegraphics[width=0.99\linewidth]{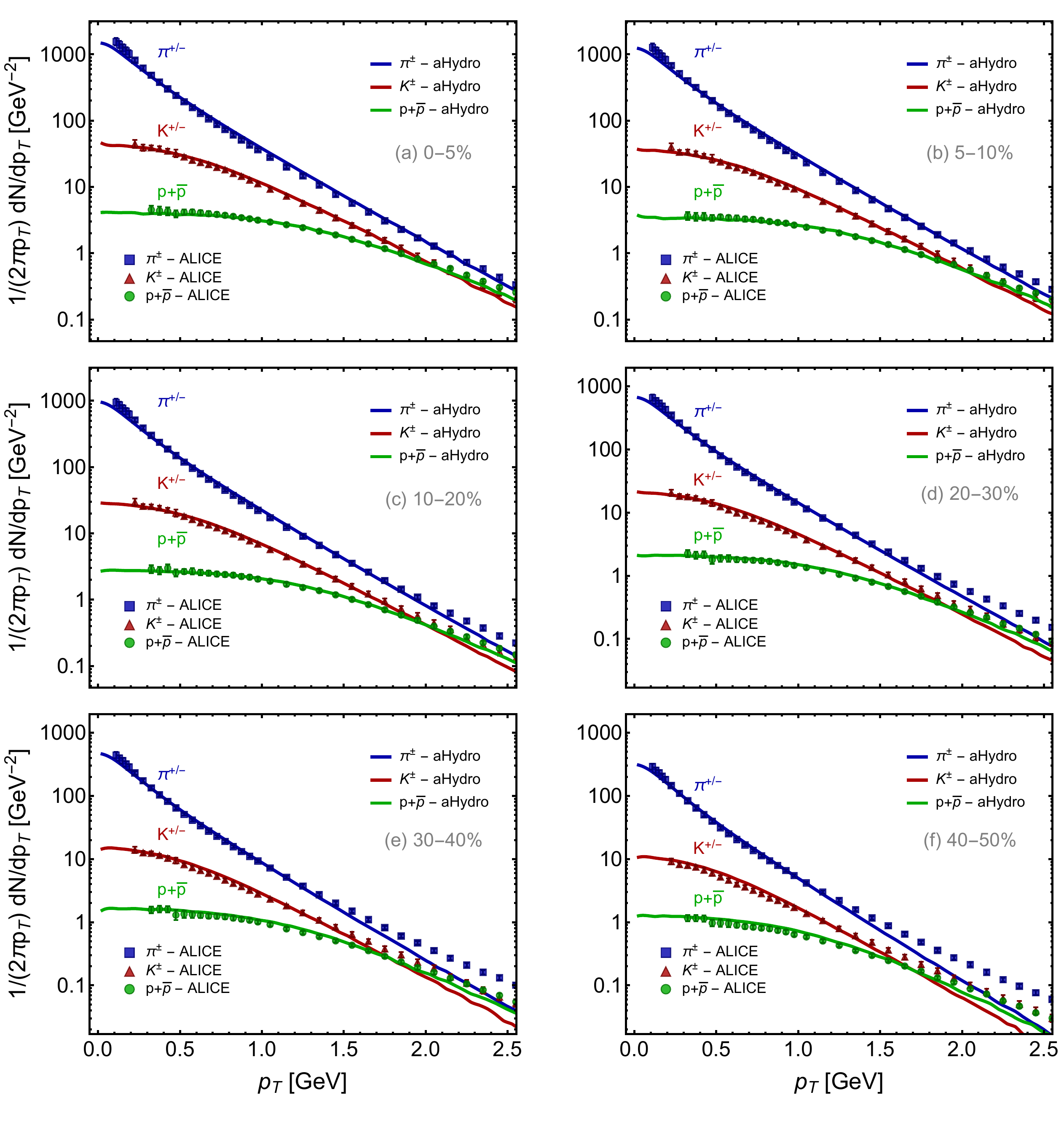}
\caption{Transverse momentum spectra of $\pi^\pm$, $K^\pm$, and $p+\bar{p}$ for six centrality classes.  All results are for 2.76 TeV Pb-Pb collisions and  data shown are from the ALICE collaboration \cite{Abelev:2013vea}.}
\label{fig:spectra}
\end{figure}

\begin{figure*}[t!]
\centerline{
\hspace{-1.5mm}
\includegraphics[width=1\linewidth]{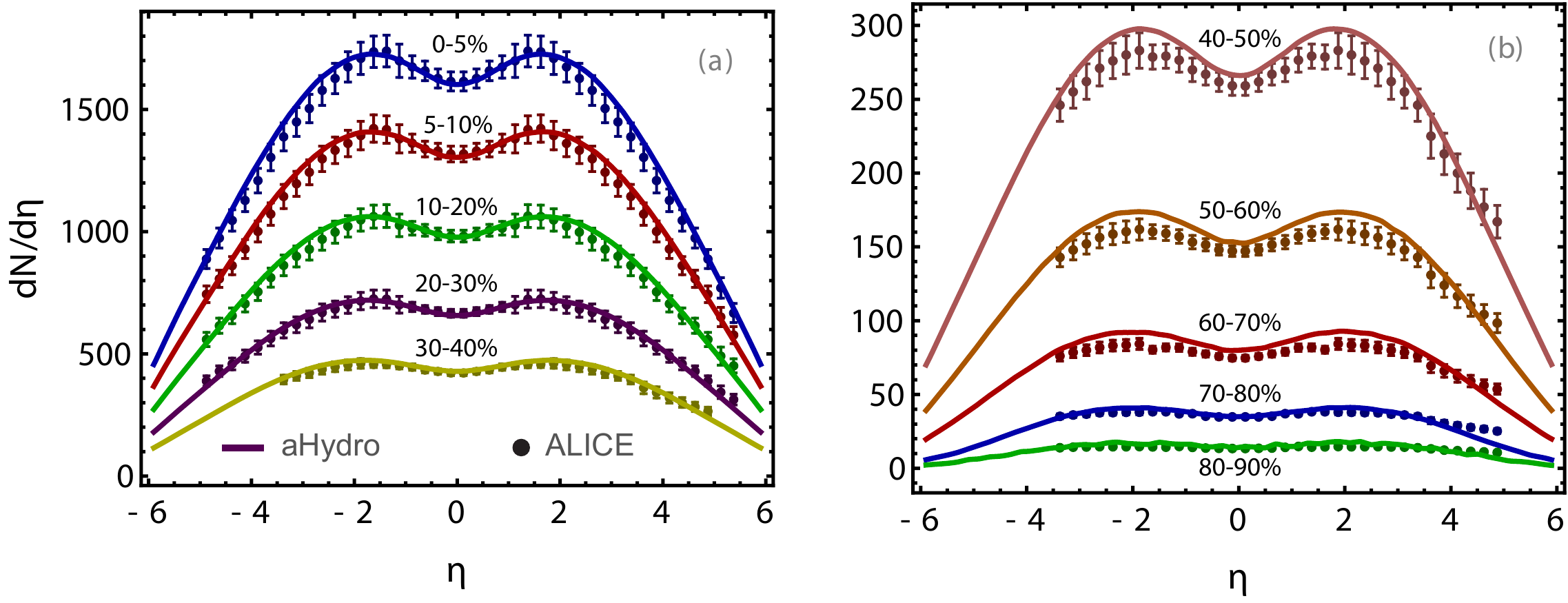}}
\caption{The charged-hadron multiplicity in different centrality classes as a function of pseudorapidity. Results are for 2.76 TeV Pb-Pb collisions and data  are from the ALICE collaboration \cite{Abbas:2013bpa,Adam:2015kda}. }
\label{fig:multiplicity}
\end{figure*}

 We solved the aHydroQP dynamical equations on a $64^3$ lattice with lattice spacings $\Delta x = \Delta y = 0.5$ fm and \mbox{$\Delta \varsigma$ = 0.375}. To compute spatial derivatives we used fourth-order centered-differences and, for temporal updates, we used fourth-order Runge-Kutta with step size of $\Delta\tau = 0.02$ fm/c. To regulate potential numerical instabilities associated with the centered-differences scheme, we used a weighted-LAX smoother~\cite{Martinez:2012tu}. In most cases, we set the weighted-LAX fraction to be $0.005$, however, for large impact parameters we used $0.02$.\footnote{This does not affect the evolution considerably since, for high impact parameters, the system reaches $T_{\rm FO}$ at times $\lesssim 4$ fm/c.} The aHydroQP evolution was started at $\tau_0 = 0.25$ fm/c and stopped when the highest effective temperature in the entire volume was sufficiently below $T_{\rm FO}$.

\begin{figure*}[t!]
\centerline{
\includegraphics[width=.5\linewidth]{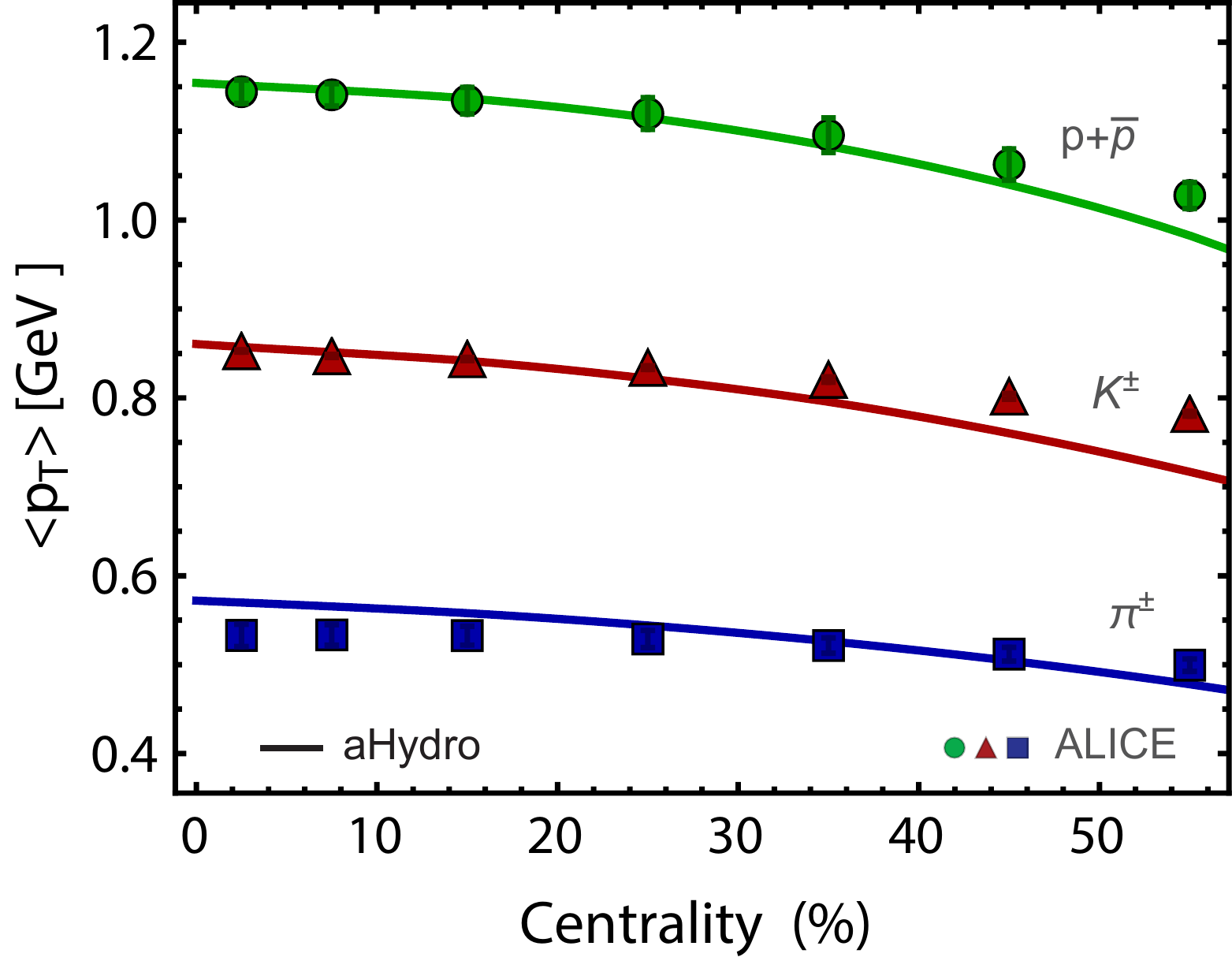}
}
\caption{The average transverse momentum of $\pi^\pm$, $K^\pm$, and $p+\bar{p}$ as a function of centrality  for 2.76 TeV Pb-Pb collisions.  Data are from the ALICE collaboration \cite{Abelev:2013vea}. }
\label{fig:ptavg}
\end{figure*}

\begin{figure*}[t!]
\centerline{\includegraphics[width=.5\linewidth]{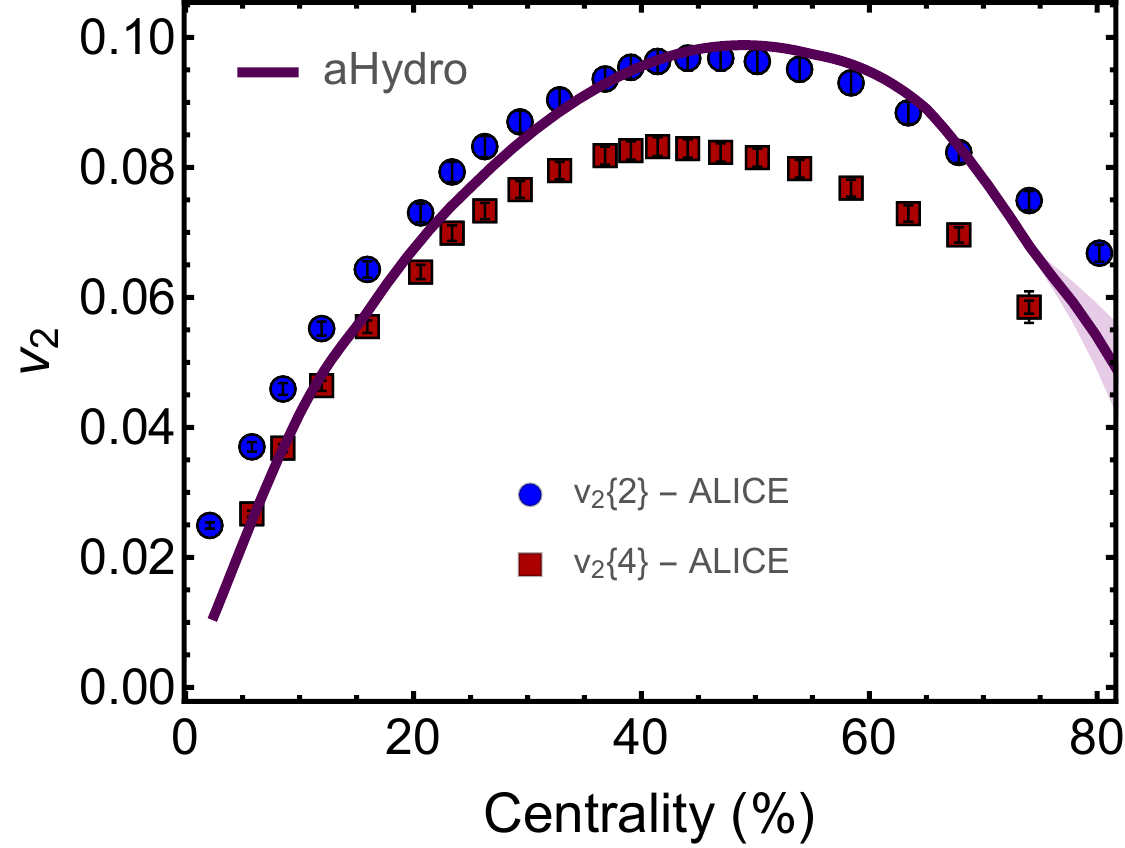}}
\caption{The integrated $v_2$ for charged hadrons as a function of centrality ($0.2 < p_T < 3$ GeV, $\eta < 0.8$).  All results are for 2.76 TeV Pb-Pb collisions and  data are from the ALICE collaboration \cite{Abelev:2014mda}. }
\label{fig:v2integrated}
\end{figure*}

\begin{figure}[t!]
\includegraphics[width=0.99\linewidth]{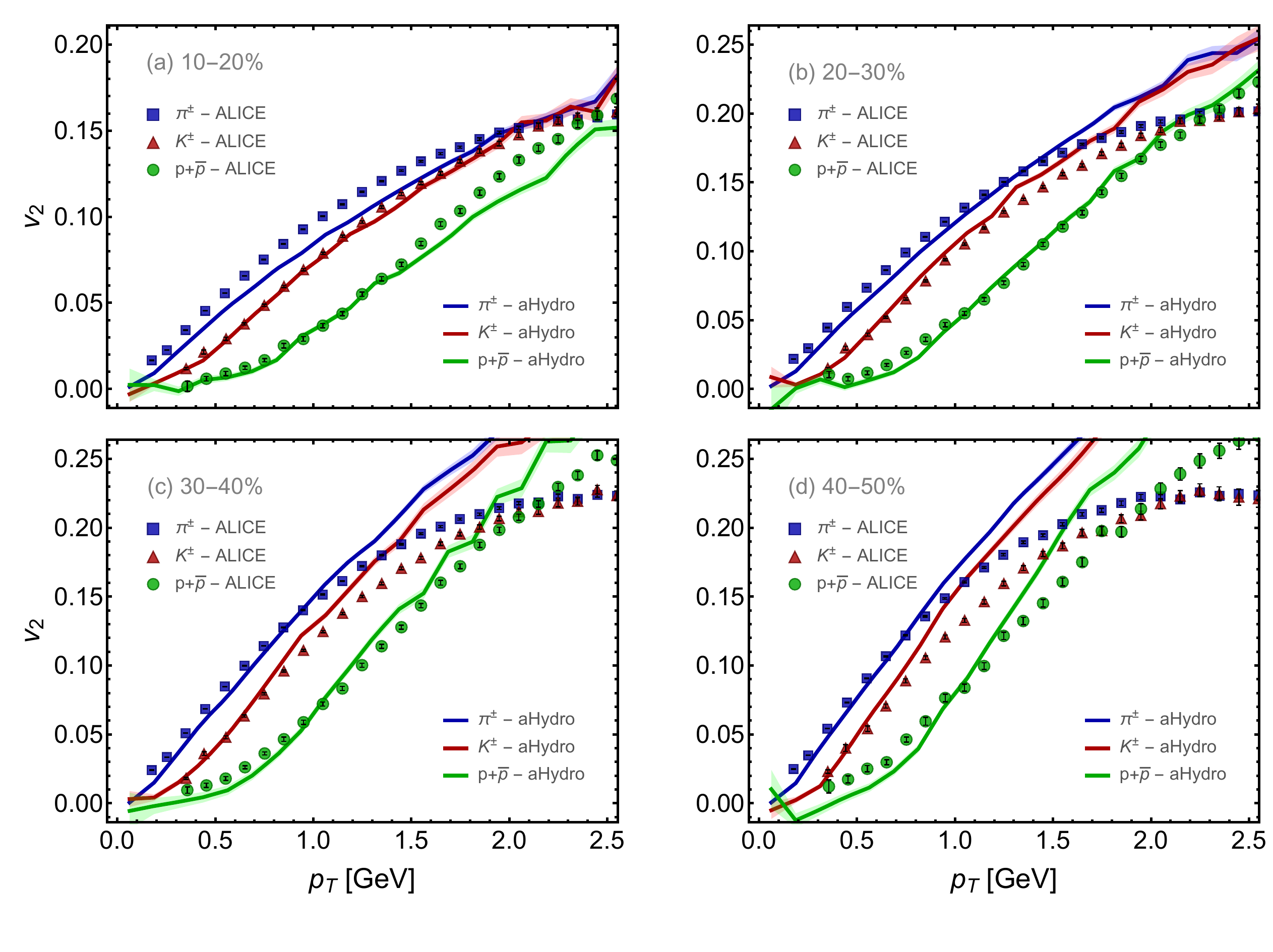}
\caption{The elliptic flow coefficient for identified hadrons as a function of  $p_T$ for four centrality classes as shown in each panel. All results and data are for 2.76 TeV Pb-Pb collisions.  Data shown are from the ALICE collaboration and were extracted using the scalar product method~\cite{Abelev:2014pua}.}
\label{fig:v2}
\end{figure}

Using aHydroQP, we first ran the full 3+1d evolution of the system, then we extracted a freeze-out hypersurface based on the effective temperature. We assumed that all hadronic species were in chemical equilibrium and had the same fluid anisotropy tensor ($\Xi_{\mu\nu}$) and scale parameter ($ \lambda$). The distribution function parameters on the freeze-out hypersurface were fed  into a customized version of THERMINATOR 2 which allows for an ellipsoidal distribution function of the form given in Eq.~(\ref{eq:genf2}). THERMINATOR 2 performs sampled event-by-event hadronic production from the exported freeze-out hypersurface  using Monte-Carlo sampling. It then performs hadronic feed down (resonance decays) for each sampled event. Depending on the observables under consideration and the centrality class considered, one may need to generate more hadronic events for the purposes of improved statistics.  For all plots shown herein we used between 7,400 and 36,200 hadronic events per centrality class. We indicate the statistical uncertainty of our model results associated with the hadronic Monte-Carlo sampling by a shaded band surrounding the hadronic event-averaged value (the central line).

In our model we have three remaining free parameters: (1) the initial central temperature $T_0$ obtained in a perfectly central collision at ${\bf x}_\perp=0$ and $\varsigma=0$, (2) the freeze-out temperature $T_{\rm FO}$, and (3) $\eta/s$ which is assumed to be a (temperature-independent) constant. In order to fix these parameters we scanned over them and compared the theoretical predictions resulting from this scan with experimental data from the ALICE collaboration for the differential spectra of pions, kaons, and protons in both the 0-5\% and 30-40\% centrality classes. The fitting error was minimized across species, with equal weighting for the three particle types.  The parameters obtained from this procedure are $T_0 = 600$ MeV, $\eta/s = 0.159$, and \mbox{$T_{\rm FO} = 130$ MeV}. 
   
\begin{figure*}[t!]
\centerline{
\hspace{-1.5mm}
\includegraphics[width=.99\linewidth]{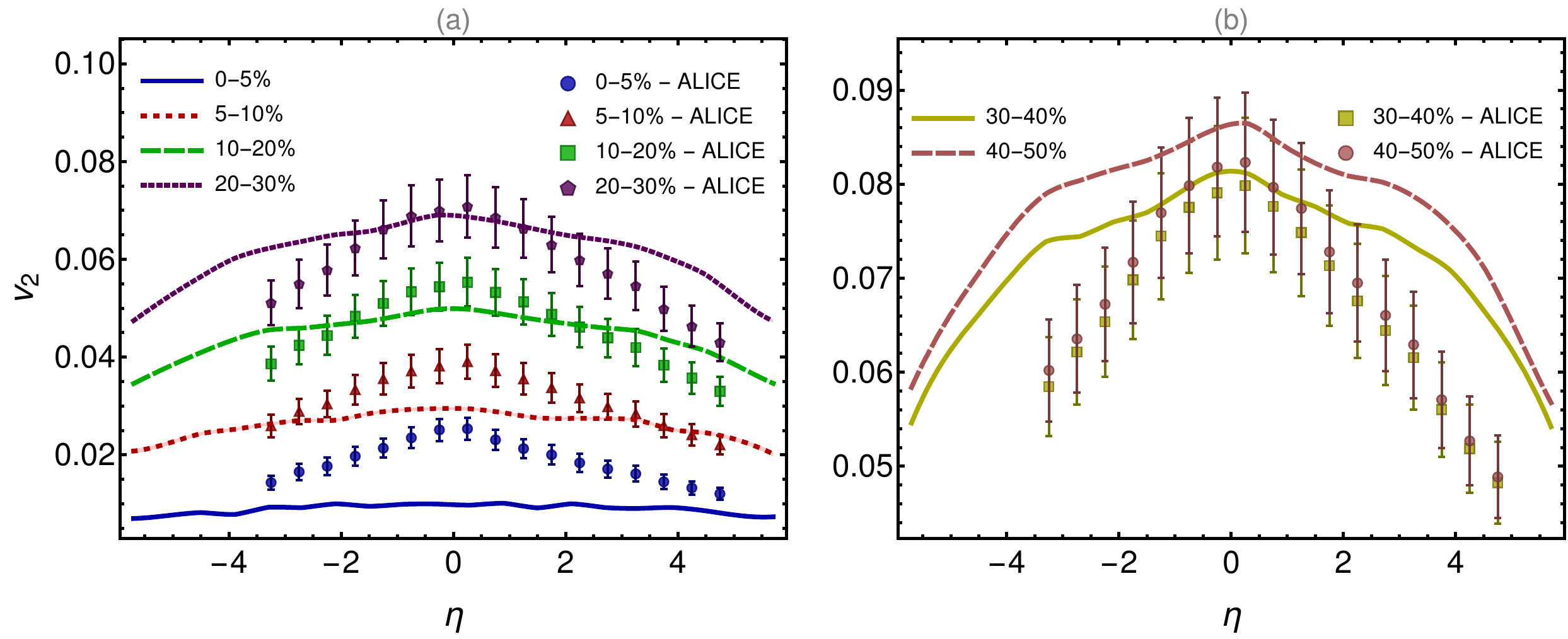}
}
\caption{The pseudorapidity dependence of the elliptic  flow $v_2$ for charged hadrons in different centrality classes where we take $0 < p_T < 100$ GeV.  All results are for 2.76 TeV Pb-Pb collisions and  data are from the ALICE collaboration \cite{Adam:2016ows}. }
\label{fig:v2_eta}
\end{figure*}

We first present our comparisons of the  transverse momentum spectra of $\pi^\pm$, $K^\pm$, and $p+\bar{p}$ in six centrality classes  0-5$\%$, 5-10$\%$, 10-20$\%$, 20-30$\%$, 30-40$\%$, and 40-50$\%$ in Fig.~\ref{fig:spectra}. These comparisons show that our model provides a very good simultaneous description of the ALICE data for the pion, kaon, and proton spectra \cite{Abelev:2013vea}, with largest differences at $p_T \gtrsim 1.5$ GeV and relatively high centrality classes 30-40$\%$, and 40-50$\%$. We note that our model slightly underpredicts the pion spectrum at low transverse momentum which is similar to what is observed in other hydrodynamic models (see e.g. Ref.~\cite{Ryu:2015vwa}). One possible explanation for this discrepancy that has been suggested is pion condensation \cite{Begun:2013nga}.

In Fig.~\ref{fig:multiplicity},  we show the charged-hadron multiplicity in different centrality classes as a function of pseudorapidity, $\eta$. In panel (a), we show the 0-5$\%$, 5-10$\%$, 10-20$\%$, 20-30$\%$, and 30-40$\%$ centrality classes, and in panel (b) we show the 40-50$\%$, 50-60$\%$, 70-80$\%$, 80-90$\%$, and  90-100$\%$ centrality classes.  As can be seen from both panels, our model is able to describe the charged hadron multiplicity as a function of pseudorapidity \cite{Abbas:2013bpa,Adam:2015kda} quite well in all centrality classes. Another observable to consider is  the average transverse momentum of pions, kaons, and protons as a function of centrality. This is shown in Fig.~\ref{fig:ptavg}, where our model is again able to reproduce the data reasonably well. 

Next, in Fig.~\ref{fig:v2integrated}, we show  the integrated elliptic flow coefficient $v_2$ for charged hadrons as a function of centrality.  Our model predictions were computed using the geometrical definition of the elliptic flow coefficient, $v_2 \sim \langle \cos(2\phi) \rangle$, for all charged hadrons.  The experimental data were obtained using second- and fourth-order cumulants $v_2\{2\}$ and $v_2\{4\}$ \cite{Abelev:2014mda}. From this figure we see that our model agrees well with $v_2\{4\}$ measurements at low centrality, but agrees better with $v_2\{2\}$ at higher centrality.  One would expect better agreement with $v_2\{4\}$ than $v_2\{2\}$, since the former has non-flow effects subtracted.  The fact that we agree better with $v_2\{2\}$ at high centrality could be due to the fact that our smooth initial condition is too simple or that we have not included the off-diagonal components of the anisotropy tensor in the evolution and freeze-out.

\begin{figure}[t!]
\centerline{
\hspace{-1.5mm}
\includegraphics[width=1\linewidth]{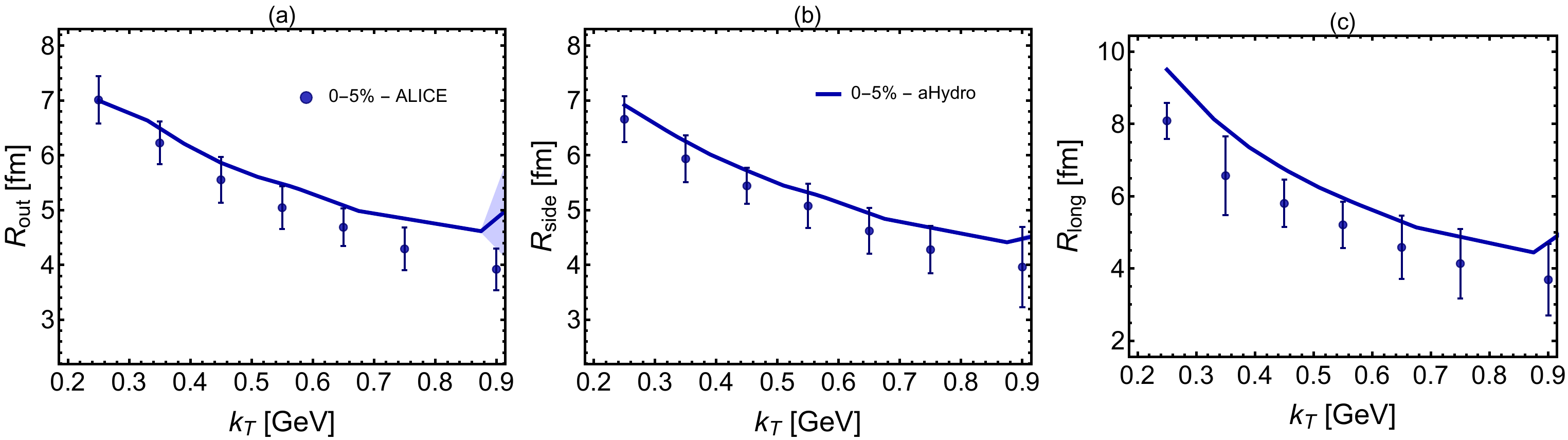}}
\centerline{
\hspace{-1.5mm}
\includegraphics[width=1\linewidth]{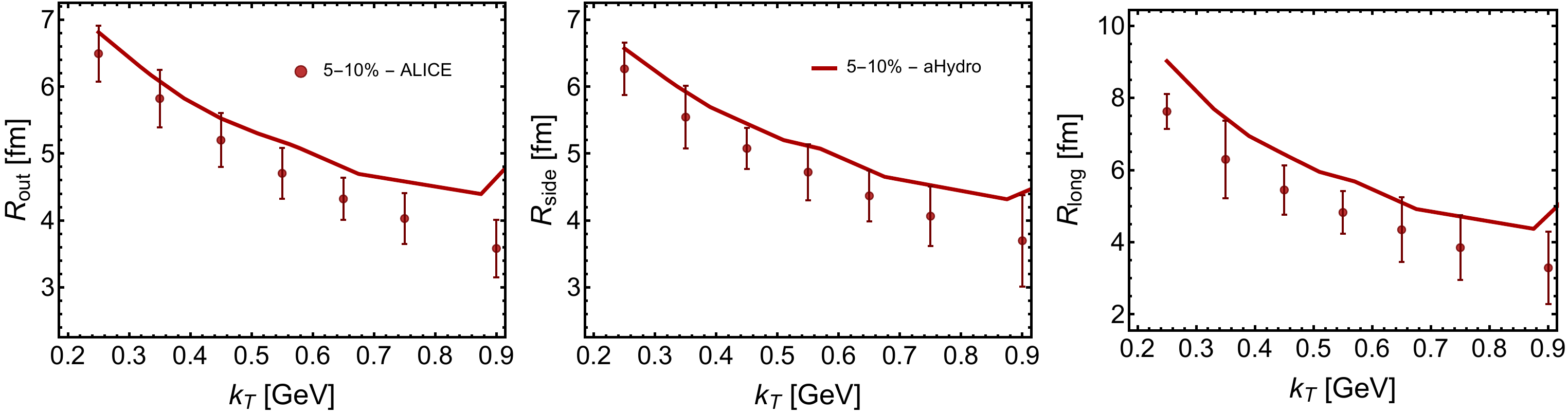}}
\centerline{
\hspace{-1.5mm}
\includegraphics[width=1\linewidth]{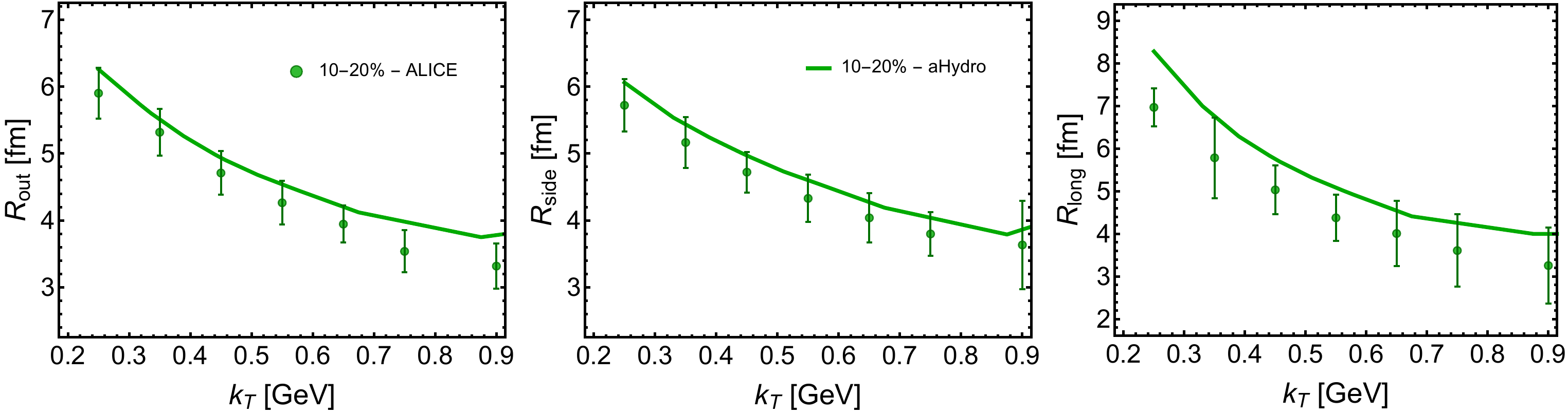}}
\centerline{
\hspace{-1.5mm}
\includegraphics[width=1\linewidth]{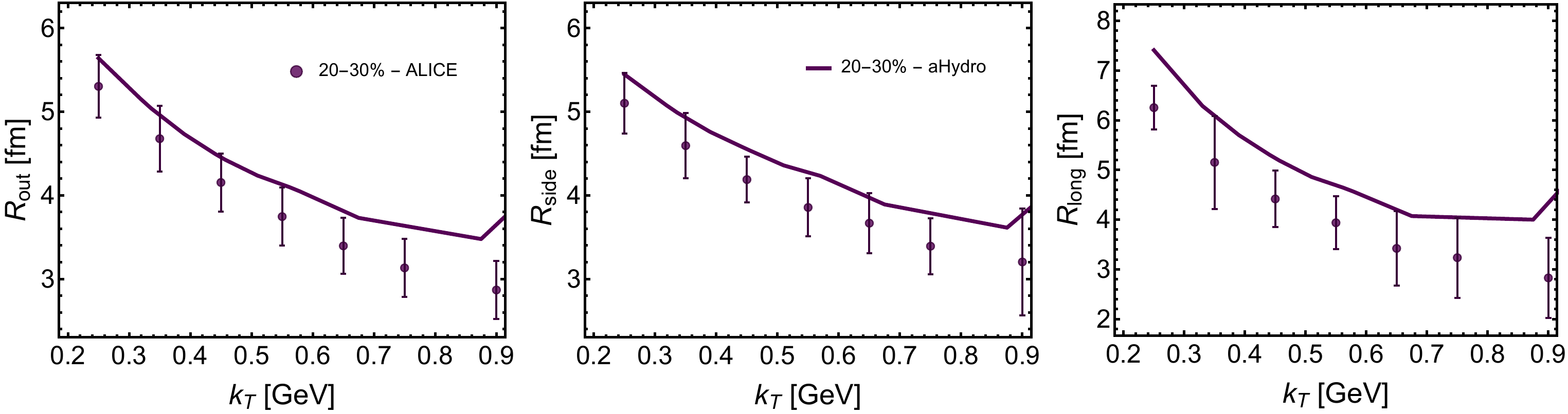}}
\caption{ The femtoscopic (HBT) radii as a function of the pair mean transverse  momentum $(k_T)$ for $\pi^+ \pi^+ $ in the  0-5$\%$, 5-10$\%$, 10-20$\%$, and 20-30$\%$ centrality classes. The left, middle, and right panels show $R_{\rm out} $, $ R_{\rm side} $, and $R_{\rm long}$, respectively. All results are for 2.76 TeV Pb-Pb collisions where  data shown are for $\pi^\pm \, \pi^\pm $ obtained by the ALICE collaboration \cite{Graczykowski:2014hoa}. }
\label{fig:HBT}
\end{figure}

In Fig.~\ref{fig:v2}, we present comparisons of the identified-particle $v_2$ as a function of $p_T$ obtained using our model with experimental data reported by ALICE collaboration~\cite{Abelev:2014pua}.  Our model provides a quite reasonable description of the identified-particle elliptic flow as can be seen in panels (b) and (c), 20-30\% and 30-40\% centrality classes, respectively.  In panel (b), the 20-30\% centrality class, we see that our model reproduces the data very well for the pion, kaon, and proton data out to $p_T \sim$ 1.5, 1.5, and 2.5 GeV, respectively.  A very similar agreement is seen in panel (c), the 30-40\% centrality class, where the model is in good  agreement with the pion, kaon, and proton data out to $p_T \sim$ 1, 1, and 2 GeV, respectively. However, in panels (a) and (d), 10-20\% and 40-50\% centrality classes, respectively, we see less agreement than panels (b) and (c). For example, we underpredict the pion elliptic flow in the 10-20$\%$ centrality class as can be seen from panel (a).  Again, as in the case of Fig.~\ref{fig:v2integrated}, this is related to our use of smooth Glauber initial conditions.

In order to further  examine how well our model describes various observables, we look at the  pseudorapidity dependence of $v_2$ for different centrality classes in Fig.~\ref{fig:v2_eta}. As can be seen from Fig.~\ref{fig:v2_eta} our model results do not fall fast enough at large pseudorapidity compared to the experimental data~\cite{Adam:2016ows}.  One possible remedy for this may be including temperature-dependent $\eta/s$, since this has been shown to improve agreement with this observable in the context of viscous hydrodynamics \cite{Denicol:2015nhu}.  
\begin{figure}[t!]
\centerline{
\hspace{-1.5mm}
\includegraphics[width=1\linewidth]{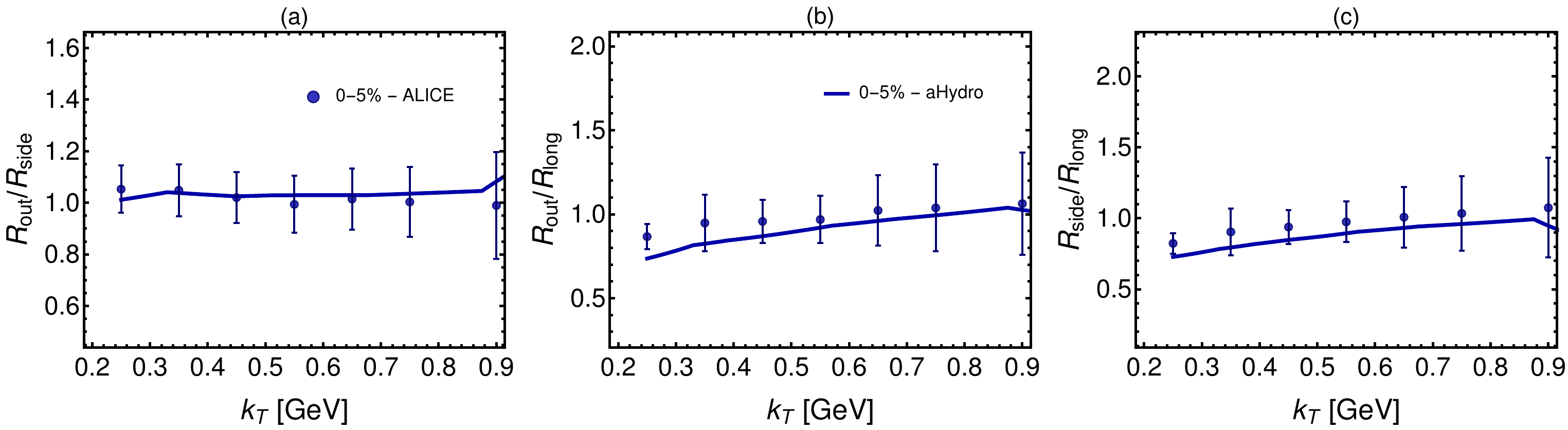}}
\centerline{
\hspace{-1.5mm}
\includegraphics[width=1\linewidth]{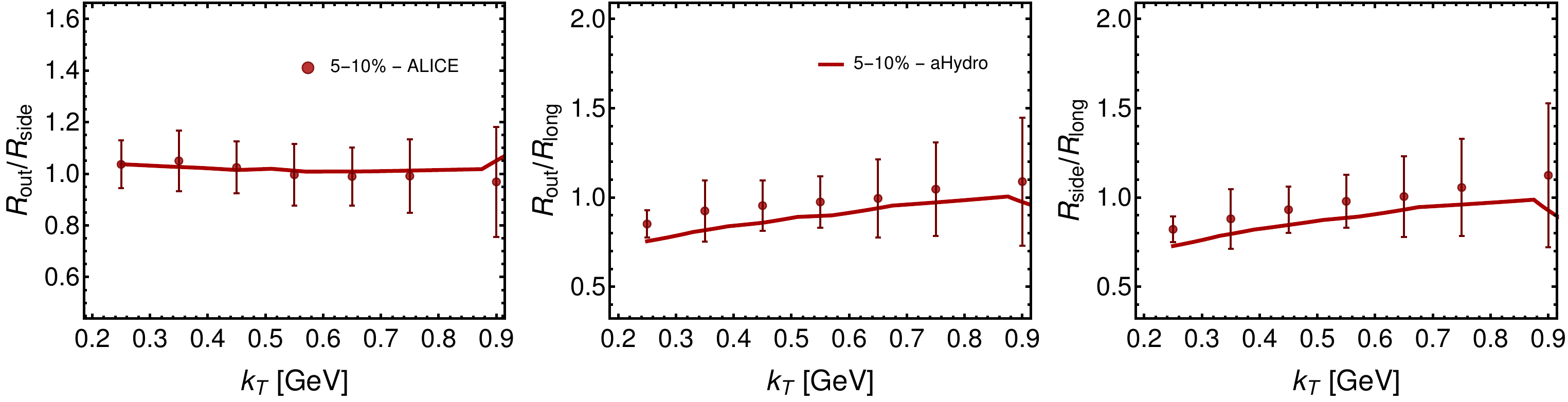}}
\centerline{
\hspace{-1.5mm}
\includegraphics[width=1\linewidth]{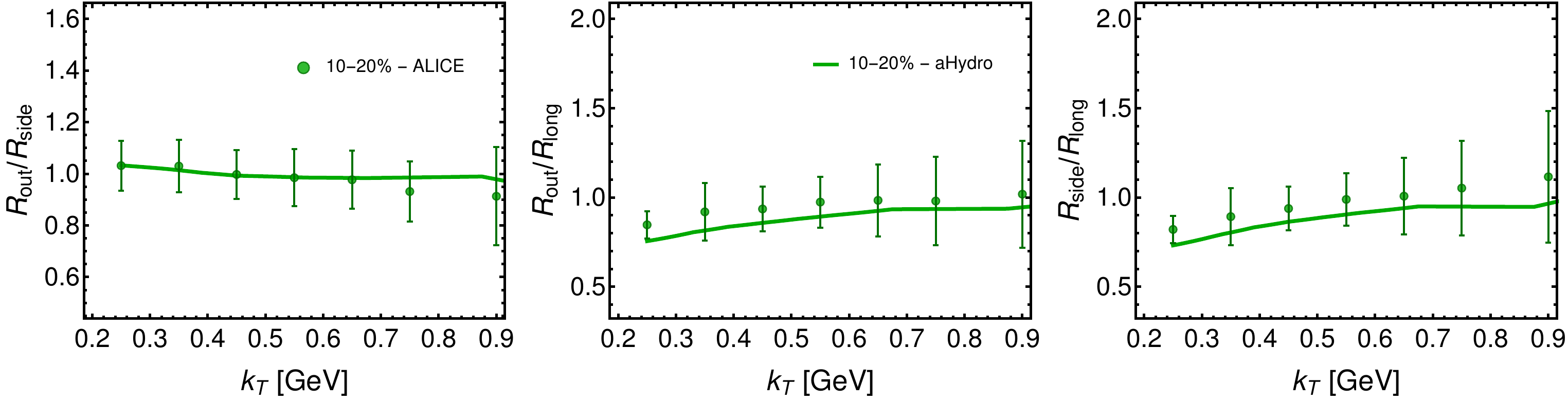}}
\centerline{
\hspace{-1.5mm}
\includegraphics[width=1\linewidth]{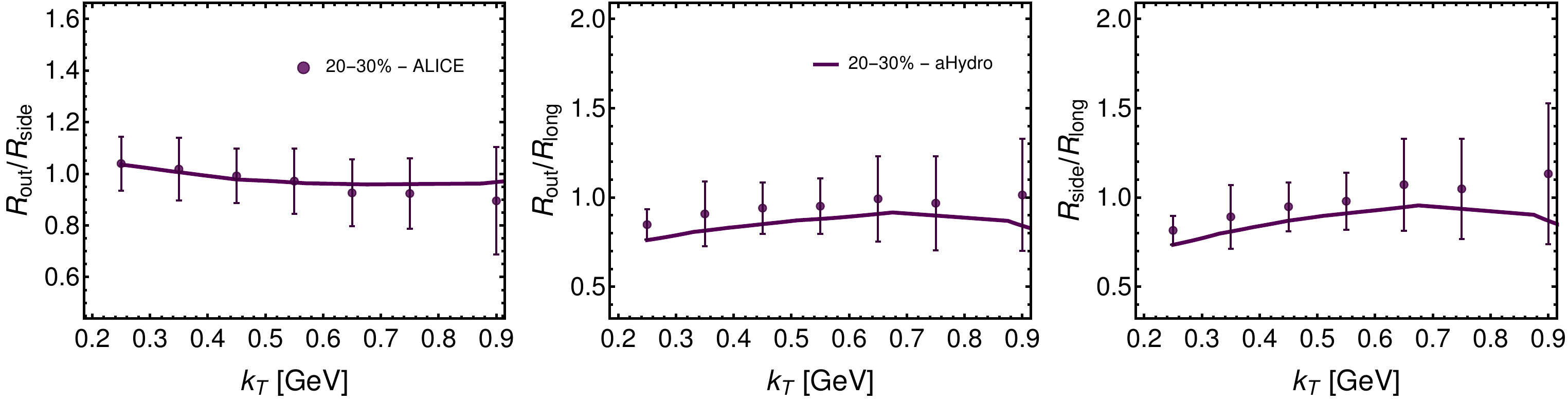}}
\caption{ The ratios of femtoscopic (HBT) radii as a function of the pair mean transverse  momentum $(k_T)$ for $\pi^+ \pi^+ $ in the  0-5$\%$, 5-10$\%$, 10-20$\%$, and 20-30$\%$ centrality classes. The left, middle, and right panels show $R_{\rm out}/R_{\rm side} $, $ R_{\rm out}/R_{\rm long} $, and $R_{\rm side}/R_{\rm long}$, respectively. All results are for 2.76 TeV Pb-Pb collisions where  data shown are for $\pi^\pm \, \pi^\pm $ obtained by the ALICE collaboration \cite{Graczykowski:2014hoa}. }
\label{fig:HBTratios}
\end{figure}

In Fig.~\ref{fig:HBT}  we  compare the HBT radii predicted by aHydroQP  with experimental data from the ALICE collaboration~\cite{Graczykowski:2014hoa}. To compute the HBT radii, we used exactly the same parameters used to describe other observables including the same number of hadronic events in each centrality class.  In this set of figures, in the left, middle, and right panels we show $R_{\rm out} $, $ R_{\rm side} $, and $R_{\rm long} $, respectively, as a function of the mean transverse momentum of the pair $\pi^+ \pi^+$ in four different centrality classes, 0-5\%, 5-10\%, 10-20\%, and 20-30\%.  From the left column of this set of figures, we see that our model reproduces the data quite well for $R_{\rm out}$ out to $k_T \sim$ 0.6 GeV. In the middle column, we present  comparisons of $R_{\rm side}$ where our model shows a good agreement out to $k_T \sim$ 0.9 GeV. Lastly, in the right column we  compare results for $R_{\rm long}$ which show poorer agreement  with the data when compared to the first two columns. However, in most cases, our model predictions are within the error bars of the experimental data, with the biggest differences at low $k_T \sim$ 0.2 GeV. This is  opposite to what we see in $R_{\rm out}$ and $R_{\rm side}$ where we observe  good agreement with the experimental data at low $k_T$. 

For more comparisons, we compare also the ratios of the HBT radii in Fig.~\ref{fig:HBTratios}. In this set of figures, in the left, middle, and right panels we show $R_{\rm out}/R_{\rm side} $, $ R_{\rm out}/R_{\rm long}  $, and $R_{\rm side}/R_{\rm long} $, respectively, as a function of the mean transverse momentum of the pair $\pi^+ \pi^+$ in four different centrality classes, 0-5\%, 5-10\%, 10-20\%, and 20-30\%. We see that our model was able to reproduce the data quite well in all three panels for all centrality classes shown here.

\section{Conclusions and Outlook}
\label{sec:conclusions}

In this paper we presented phenomenological comparisons of aHydroQP with LHC experimental data collected in 2.76 TeV Pb-Pb collisions. This work is an extension of a previous letter \cite{Alqahtani:2017jwl}.  Herein, we gave more details about the formalism used and presented a more thorough comparison between our model and LHC data for a variety of observables. In  aHydroQP, we included three momentum anisotropy parameters in the underlying distribution function, both in the dissipative hydrodynamic stage and at freeze-out. We also used a quasiparticle implementation of the LQCD EoS in order to take into account the non-conformality of the system.  At freeze-out, we used a customized version of THERMINATOR 2 which was modified to accept anisotropic distribution functions of generalized Romatschke-Strickland form.  As a first test, in this work, we used smooth Glauber initial conditions which were obtained from  a linear combination of wounded-nucleon and binary-collision profiles. We additionally assumed the system to be initially isotropic in momentum space with no initial transverse flow. 

To fix the remaining phenomenological parameters, we performed a parameter scan and compared our results with experimentally observed identified-particle spectra in the 5-10\% and 30-40\% centrality classes. The resulting set of best fit parameters was $T_0 = 600$ MeV, $\eta/s = 0.159$, and \mbox{$T_{\rm FO} = 130$ MeV}.  After this fitting was complete, we computed an array of different heavy-ion observables, finding quite good agreement between our model and experimental data despite our simple smooth initial condition. We looked at particle multiplicity and spectra, average transverse momentum, $v_2$, and HBT radii.  Compared to Ref.~\cite{Alqahtani:2017jwl}, we have added additional centrality classes in some cases and increased the statistics associated with the hadronic Monte-Carlo sampling where necessary.

Combined with what was reported in \cite{Alqahtani:2017jwl}, the phenomenological results presented herein represent the first aHydro results to include three separate anisotropy parameters together with the quasiparticle method for imposing the EoS and self-consistent anisotropic freeze-out.  Compared to prior results which used a single anisotropy parameter and/or an approximate conformal-factorization implementation of the equation of state \cite{Ryblewski:2012rr,Nopoush:2016qas,Strickland:2016ezq} we see much better agreement with the pion, kaon, and proton spectra and, relatedly, the total multiplicity as a function of pseudorapidity.  Prior studies which used the approximate conformal-factorization implementation of the equation of state dramatically underestimated the low $p_T$ spectra \cite{Nopoush:2016qas,Strickland:2016ezq}, making this the first phenomenological study within the context of aHydro which is able to reproduce both the experimentally observed spectra and elliptic flow.

Looking to future, there is certainly room for improvements in our model. For example, we are working on including a temperature-dependent shear viscosity to entropy density ratio since in this study it was assumed to be constant. Based on prior studies in the context of viscous hydrodynamics~\cite{Denicol:2015nhu}, there is some hope that this will improve the agreement between our model and the experimental data, in particular, with regards to the pseudorapidity dependence of $v_2$. We also plan to include realistic fluctuating initial conditions, realistic initial momentum anisotropy profiles, and more realistic collisional kernels. Additionally, we are planning to look at different collision energies, e.g.  RHIC 200 GeV collisions and LHC 5.023 TeV collisions, and different colliding systems, e.g. pA and pp, in the near future.  The application of aHydro to pA and pp is of particular interest, since in these systems viscous hydrodynamics is being pushed to its limits, especially at freeze-out \cite{Nopoush:2015yga}.  

Finally, it would be interesting to apply this formalism also to even lower energy collisions where it is critical to take into account the finite net baryon density, heat flow, etc.  In this context, it would also be interesting to extend the formalism to multicomponent fluids, e.g. two- and three-fluid models, similar to what has been done by the Los Alamos \cite{Amsden:1978zz,Clare:1986qj}, Kurchatov Institute \cite{Mishustin:1988mj,Mishustin:1991sp,Ivanov:2000dr}, Frankfurt \cite{Katscher:1993xs,Dumitru:1994vc,Brachmann:1997bq,Reiter:1998uq,Brachmann:1999xt}, and GSI \cite{Russkikh:1993ct,Ivanov:1991te,Ivanov:1993ng,Ivanov:1995wf,Russkikh:1995yz,Ayik:1994zi} groups.  Along these lines, there have already been prior anisotropic hydrodynamics studies which have considered multi-component fluids with explicit quark and gluon components \cite{Florkowski:2012as,Florkowski:2015cba}; however, to the best of our knowledge there have thus far not been any attempts to do this in the context of multicomponent hadronic fluids.

\acknowledgments{ We thank Piotr Bo\.zek for useful comments. M.~Alqahtani was supported by a PhD fellowship from the Imam Abdulrahman Bin Faisal University, Saudi Arabia.  M.~Nopoush and M.~Strickland were supported by the U.S. Department of Energy, Office of Science, Office of Nuclear Physics under Award No. DE-SC0013470.  R.~Ryblewski was supported by the Polish National Science Center grant No. DEC-2012/07/D/ST2/02125.}

\appendix

\section{Explicit formulas for derivatives}
\label{app:identities}

In this section, we introduce the notation used in our formulation of the general moment-based hydrodynamics equations. Using the definitions
\ba
{\cal D}&\equiv&\cosh(\vartheta-\varsigma)\partial_\tau+\frac{1}{\tau}\sinh(\vartheta-\varsigma)\partial_\varsigma\nonumber\,, \\
\tilde{{\cal D}}&\equiv&\sinh(\vartheta-\varsigma)\partial_\tau+\frac{1}{\tau}\cosh(\vartheta-\varsigma)\partial_\varsigma\,, \\
\nabla_\perp\cdot{\bf u}_\perp &\equiv&\partial_x u_x+\partial_y u_y \nonumber\,, \\
{\bf u}_\perp\cdot\nabla_\perp &\equiv& u_x\partial_x+u_y\partial_y \nonumber\,,\\
{\bf u}_\perp\times\nabla_\perp &\equiv& u_x \partial_y-u_y\partial_x\,,
\label{eq:identities-gen}
\ea
and four-vectors defined in Eq.~(\ref{eq:4vectors}) one obtains
\ba
D_u&\equiv&u^\mu \partial_\mu=u_0{\cal D}+{\bf u}_\perp\cdot\nabla_\perp \nonumber\,,\\
D_x&\equiv&X^\mu \partial_\mu=u_\perp{\cal D}+\frac{u_0}{u_\perp}({\bf u}_\perp\cdot\nabla_\perp)\nonumber\,,\\
D_y&\equiv&Y^\mu \partial_\mu=\frac{1}{u_\perp}({\bf u}_\perp\times\nabla_\perp)\nonumber\,,\\
D_z&\equiv&Z^\mu \partial_\mu=\tilde{{\cal D}}\,.
\ea
The divergences can be defined as
\ba 
\theta_u&\equiv&\partial_\mu u^\mu={\cal D}u_0+u_0\tilde{{\cal D}}\vartheta+\nabla_\perp\cdot{\bf u}_\perp\nonumber\,,\\
\theta_x&\equiv&\partial_\mu X^\mu={\cal D}u_\perp+u_\perp\tilde{{\cal D}}\vartheta+\frac{u_0}{u_\perp}(\nabla_\perp\cdot{\bf u}_\perp)-\frac{1}{u_0 u_\perp^2}({\bf u}_\perp\cdot\nabla_\perp) u_\perp\nonumber\,,\\
\theta_y&\equiv&\partial_\mu Y^\mu=-\frac{1}{u_\perp}({\bf u}_\perp\cdot\nabla_\perp) \varphi\nonumber\,, \\
\theta_z&\equiv&\partial_\mu Z^\mu={\cal D}\vartheta\,,
\label{eq:deriv-gen}
\ea
where $\varphi=\tan^{-1}(u_y/u_x)$. Finally, we list the non-vanishing contractions appearing in the second moment equations  
\ba
u_\mu D_\alpha X^\mu &=&\frac{1}{u_0}D_\alpha u_\perp \nonumber\,, \\
u_\mu D_\alpha Y^\mu &=& u_\perp D_\alpha\varphi \nonumber \,, \\
u_\mu D_\alpha Z^\mu &=& u_0 D_\alpha\vartheta \nonumber \,,\\
X_\mu D_\alpha Y^\mu &=& u_0 D_\alpha\varphi \nonumber \,,\\
X_\mu D_\alpha Z^\mu &=& u_\perp D_\alpha\vartheta \nonumber\,, \\
Y_\mu D_\alpha Z^\mu &=& 0\,, \label{eq:iden2-gen}
\ea
where $\alpha\in\{u,x,y,z\}$. Note that, using the orthogonality of the basis vectors, i.e. $D_\alpha(X^\mu u_\mu) = 0$, contractions such as $X^\mu D_\alpha u_\mu$ can be related to the contractions listed above, e.g.   $X^\mu D_\alpha u_\mu = -u_\mu D_\alpha X^\mu$. 

\section{Special functions and derivatives}
\label{app:h-functions}

The $ {\cal H}$-functions appearing in the body of the paper can be written as
\ba 
{\cal H}_3({\boldsymbol\alpha},\hat{m}) &\equiv&\tilde{N} \alpha \int d^3\hat{p} \, {\cal R}  \, f_{\rm eq}\!\left(\!\sqrt{\hat{p}^2 + \hat{m}^2}\right) , \\
{\cal H}_{3i}({\boldsymbol\alpha},\hat{m}) &\equiv& \tilde{N} \alpha \, \alpha_i^2 \int d^3\hat{p} \,{\cal R}_i  \, f_{\rm eq}\!\left(\!\sqrt{\hat{p}^2 + \hat{m}^2}\right) , \\
{\cal H}_{3B}({\boldsymbol\alpha},\hat{m}) &\equiv& \tilde{N} \alpha \int d^3\hat{p} \, {\cal R}^{-1}  \, f_{\rm eq}\!\left(\!\sqrt{\hat{p}^2 + \hat{m}^2}\right) , 
\ea
where $\boldsymbol\alpha = (\alpha_x,\alpha_y,\alpha_z)$, $\hat{m}=m/\lambda$, $i \in \{x,y,z\}$, $\alpha \equiv \prod_i \alpha_i$, and $ \tilde{N}\equiv N_{\rm dof}/(2\pi)^3$ with $ N_{\rm dof}$ being the number of degrees of freedom.  The ${\cal R}$ and ${\cal R}_i$ functions appearing above are
\ba 
{\cal R} &\equiv& \sqrt{\alpha_x^2 \,  \hat{p}_x^2+\alpha_y^2 \,  \hat{p}_y^2+\alpha_z^2 \,   \hat{p}_z^2+\hat{m}^2} \, , \\
{\cal R}_i &\equiv& \hat{p}_i^2{\cal R}^{-1} \, .
\ea
More details concerning the ${\cal H}$-functions and the manner in which they appear in the dynamical equations can be found in Refs.~\cite{Alqahtani:2015qja} and \cite{Alqahtani:2016rth}.

To the best of our knowledge, it is not possible to analytically evaluate the ${\cal H}$-functions listed above. In practice, only one integral can be done analytically and we are left with integrals over $\phi $ and $p$. Evaluation of these 2d integrals is numerically intensive, making it infeasible to evaluate them in real-time during 3+1d simulations.  One might consider interpolating them, however, they are functions of 4 variables $ ({\boldsymbol\alpha},\hat{m})$ and, in practice, one must separately interpolate these five functions and all derivatives necessary.  As a consequence, one quickly runs into memory limitations, even on modern computers.  A more efficient technique, which does not require a great deal of memory either, is needed.  In the next subsection, we present a method for doing this.

\subsection*{Series expansions}

Since, in practice, the $\alpha$'s do not evolve too far from $\alpha_x = \alpha_y = \alpha_z =1$, it makes sense to expand these integrals around such an isotropic point.  After expanding around an isotropic point, the angular part of the integrals become trivial and one is left only with the $p$ integral which can easily be interpolated since it is only function of the mass.  Before proceeding, we note that many of the $ {\cal H}$-functions are related to each other by symmetries. As a result, we will present the method for evaluating ${\cal H}_3$ and ${\cal H}_{3x}$ and use symmetries to find the other ${\cal H}$-functions and their derivatives necessary.

To proceed, we expand around an arbitrary isotropic point defined by $\alpha_i^2 \sim \delta_0 $ using $ \alpha_i^2 = \delta_0 + \delta_i \, \epsilon$ where $ \delta_0$ is the point around which the expansion is performed, and $\epsilon$ is used to keep track of the order of the expansion.  Based on this, we expand ${\cal R}$ and ${\cal R}_x$ as
\ba 
&&{\cal R}= \sum _{n=0}^{\infty} \, \binom{\frac{1}{2}}{n} \left(\hat{m}^2+\delta_0 \, \hat{p}^2\right)^{\frac{1}{2}-n} \hat{p}^{2n} \big[\sin ^2\theta \left(\delta_x \,
   \cos ^2\phi+\delta_y \, \sin ^2\phi\right)+\delta_z \, \cos ^2\theta \big]^n \, , \nonumber \\ 
&& {\cal R}_x= \sum _{n=0}^{\infty} \, \binom{-\frac{1}{2}}{n} \left(\hat{m}^2+\delta_0 \, \hat{p}^2\right)^{-\frac{1}{2}-n} \hat{p}^{2+2n} \cos^2 \phi \sin^2 \theta \big[\sin ^2\theta \left(\delta_x \,
   \cos ^2\phi+\delta_y \, \sin ^2\phi\right)+\delta_z \, \cos ^2\theta \big]^n \, , \nonumber
\ea
where $\delta_i=\alpha_i^2-\delta_0$.

As a result, ${\cal H}_3({\boldsymbol\alpha},\hat{m}) $ can be written as
\ba 
{\cal H}_3({\boldsymbol\alpha},\hat{m}) = \tilde{N} \alpha \sum_{n=0}^{\infty}  \binom{\frac{1}{2}}{n} \, \Omega( \boldsymbol\delta,n) {\cal G}(\frac{1}{2}-n,\hat{m},\delta_0) \, ,
\ea
where $ \Omega( \boldsymbol\delta,n) $ is the angular part of the integral which is trivial to evaluate
\be
\Omega( \boldsymbol\delta,n) = \int d\Omega \big[\sin ^2\theta \left(\delta_x \,
   \cos ^2\phi+\delta_y \, \sin ^2\phi\right)+\delta_z \, \cos ^2\theta \big]^n \, ,
\ee
and
\be 
{\cal G}(a,\hat{m},\delta_0) = \int\!d\hat{p} \, \hat{p}^{3-2 a}  \, \left(\hat{m}^2+\delta_0 \hat{p}^2\right)^a \, f_{\rm eq}\!\left(\!\sqrt{\hat{p}^2 + \hat{m}^2}\right)  . 
\ee
Similarly, $ {\cal H}_{3x}({\boldsymbol\alpha},\hat{m}) $ can be written as
\ba 
{\cal H}_{3x}({\boldsymbol\alpha},\hat{m}) = \tilde{N} \alpha \, \alpha_x^2 \sum_{n=0}^{\infty}  \binom{-\frac{1}{2}}{n} \, \Omega_x( \boldsymbol\delta,n) {\cal G}(-\frac{1}{2}-n,\hat{m},\delta_0) \, ,
\ea
where $ \Omega_x( \boldsymbol\delta,n) $ is the angular part of the integral which is trivial 
\be
\Omega_x( \boldsymbol\delta,n) = \int d\Omega \, \cos^2 \phi \sin^2 \theta \, \big[\sin ^2\theta \left(\delta_x \,
   \cos ^2\phi+\delta_y \, \sin ^2\phi\right)+\delta_z \, \cos ^2\theta \big]^n \, .
\ee
The derivatives of both $ {\cal H}_3$ and $ {\cal H}_{3x}$ with respect to $\alpha$'s are straightforward since the $ {\cal G}(a,\hat{m},\delta_0)$ integral is independent of ${\boldsymbol\alpha}$. 

There are symmetries of each ${\cal H}$-function that can be used for efficiently computing all derivatives necessary, for example, $ {\cal H}_3$ is symmetric under the exchange of $\alpha$'s, i.e.,
\ba 
{\cal H}_3(\alpha_x,\alpha_y,\alpha_z,\hat{m}) ={\cal H}_3(\alpha_y,\alpha_x,\alpha_z,\hat{m}) ={\cal H}_3(\alpha_z,\alpha_y,\alpha_x,\hat{m}) \, .
\ea 
In a similar way, 
\ba 
{\cal H}_{3x}(\alpha_x,\alpha_y,\alpha_z,\hat{m}) ={\cal H}_{3x}(\alpha_x,\alpha_z,\alpha_y,\hat{m}) \, .\label{eq:H3Xsym}
\ea 
Using these identities, once one calculates one of these derivatives, the other ones can be determined using symmetry arguments. For example, once $ \partial {\cal H}_3/\partial \alpha_x$ is known, the other derivatives with respect to $\alpha_y$ and $\alpha_z$ are related by the exchange symmetry
\ba 
\frac{\partial {\cal H}_3(\alpha_x,\alpha_y,\alpha_z,\hat{m}) }{\partial \alpha_y} = \frac{\partial {\cal H}_3(\alpha_y,\alpha_x,\alpha_z,\hat{m}) }{\partial \alpha_x}\, , \\
\frac{\partial {\cal H}_3(\alpha_x,\alpha_y,\alpha_z,\hat{m}) }{\partial \alpha_z} = \frac{\partial {\cal H}_3(\alpha_z,\alpha_y,\alpha_x,\hat{m})}{\partial \alpha_x} \, .
\ea
Unlike $ {\cal H}_3$, for $ {\cal H}_{3x}$ we have only one identity to use, so we need two derivatives,  $\partial {\cal H}_3/\partial \alpha_x$ and  $ \partial {\cal H}_3/\partial \alpha_y$ and then  Eq.~(\ref{eq:H3Xsym}) can be used to find $ \partial {\cal H}_3/\partial \alpha_z$ 
\ba 
\frac{\partial {\cal H}_{3x}(\alpha_x,\alpha_y,\alpha_z,\hat{m}) }{\partial \alpha_z} = \frac{\partial {\cal H}_{3x}(\alpha_x,\alpha_z,\alpha_y,\hat{m})}{\partial \alpha_y} \, .
\ea
We now turn to the derivative with respect to the fourth argument, $\hat{m}$. The only part in $ {\cal H}_3$ and $ {\cal H}_{3x}$ that involves $\hat{m}$ is the integral $ {\cal G}(a,\hat{m},\delta_0)$. Taking its derivative gives another integral which can be easily interpolated and used
\be
{\cal G}_{\rm {m}}(a,\hat{m},\delta_0) = - \int\!d\hat{p} \, \frac{\hat{p}^{3-2 a} \left(\hat{m}^2+\delta_0 \, \hat{p}^2\right)^{a-1}}{2 \sqrt{\hat{m}^2+\hat{p}^2}}  \left(\hat{m}^2-2 a \sqrt{\hat{m}^2+\hat{p}^2}+\delta_0 \, \hat{p}^2\right) \, f_{\rm eq}\!\left(\!\sqrt{\hat{p}^2 + \hat{m}^2}\right) . \nonumber 
\ee
So,
\ba 
\frac{\partial {\cal H}_{3}}{\partial \hat{m}} &=& 2 \hat{m} \alpha \tilde{N}  \sum_{n=0}^{\infty}  \binom{\frac{1}{2}}{n} \, \Omega( \boldsymbol\delta,n) {\cal G}_m(\frac{1}{2}-n,\hat{m},\delta_0) \, , \\
\frac{\partial {\cal H}_{3x}}{\partial \hat{m}} &=& 2 \hat{m} \alpha_x^2 \alpha  \tilde{N} \sum_{n=0}^{\infty}  \binom{-\frac{1}{2}}{n} \, \Omega_x( \boldsymbol\delta,n) {\cal G}_m(-\frac{1}{2}-n,\hat{m},\delta_0) \, .
\ea
Using the symmetries obeyed by the $ {\cal H}$ functions, one can evaluate $ {\cal H}_{3y}$, ${\cal H}_{3z}$, and ${\cal H}_{3B}$ and their derivatives similarly
\ba 
{\cal H}_{3y}(\alpha_x,\alpha_y,\alpha_z,\hat{m})  &=& {\cal H}_{3x}(\alpha_y,\alpha_x,\alpha_z,\hat{m}) \, ,   \\
{\cal H}_{3z}(\alpha_x,\alpha_y,\alpha_z,\hat{m})  &=&{\cal H}_{3x}(\alpha_z,\alpha_y,\alpha_x,\hat{m}) \, , \\
{\cal H}_{3B}({\boldsymbol\alpha},\hat{m}) &=& \frac{1}{\hat{m}^2}\big({\cal H}_{3}({\boldsymbol\alpha},\hat{m})-{\cal H}_{3x}({\boldsymbol\alpha},\hat{m})-{\cal H}_{3y}({\boldsymbol\alpha},\hat{m})-{\cal H}_{3z}({\boldsymbol\alpha},\hat{m})\big) \, .
\ea

\subsection*{Expansion points}

Finally, we must specify which value(s) of $\delta_0$ to use and the order of the expansion.  Since the $\alpha_i$'s are typically in a region $0 < \alpha_i \lesssim 3$ during the dynamical evolution, we expand around two points corresponding to $\delta_0 =1,4$ and interpolate between these two expansions in the intermediate region.  For the interpolation, we define $r_{\rm min}=1.75 $ and $r_{\rm max}=1.85 $ where $r= \sqrt{\alpha_x^2+\alpha_y^2+\alpha_z^2}$ and, for ${\cal H}_{3}$, use
\ba 
{\cal H}_{3}({\boldsymbol\alpha},\hat{m},\delta_0)=
\left\{ \begin{array}{lcc}
{\cal H}_{3}({\boldsymbol\alpha},\hat{m},1) & \mbox{If}
& r<r_{\rm min} \, ,
 \\ {\cal H}_{3}({\boldsymbol\alpha},\hat{m},4) & \mbox{If}
& r>r_{\rm max} \, , \\
\frac{r_{\rm max}-r}{r_{\rm max}-r_{\rm min}}{\cal H}_{3}({\boldsymbol\alpha},\hat{m},1)+\frac{r-r_{\rm min}}{r_{\rm max}-r_{\rm min}}{\cal H}_{3}({\boldsymbol\alpha},\hat{m},4) & 
& \mbox {otherwise} \, .
\end{array}\right. \,\,\,\,\,\,\,\,\,\,\,\,\,
\ea 
In a similar way, ${\cal H}_{3x}$ and all derivatives necessary can be calculated. In all cases, we expand up to $12^{\rm th}$ order ($n \leq 12$) which was found to reproduce the direct numerical evaluation of all ${\cal H}$-functions very well.

\bibliography{3p1ahydro}

\end{document}